\documentclass[a4paper, twocolumn]{revtex4}
\usepackage[dvips]{graphicx} 
\usepackage{amsmath}
\usepackage{amsfonts}
\usepackage{amssymb}
\usepackage{bm}

\newcommand{\be}{\begin{equation}}
\newcommand{\ee}{\end{equation}}

\newcommand{\ef}[1]{\, #1}

\newcommand{\eval}[1]{\left\langle {#1} \right\rangle}

\newcommand{\idf}{\mathrm{Id}_{\mathit{f}}}
\newcommand{\idr}{\mathrm{Id}_{\mathit{r}}}

\newcommand{\eps}{\epsilon}
\newcommand{\vx}{\boldsymbol{x}}
\newcommand{\vy}{\boldsymbol{y}}
\newcommand{\nh}{\widehat{\boldsymbol{n}}}
\newcommand{\vxi}{\boldsymbol{\xi}}

\newcommand{\vzero}{\boldsymbol{0}}
\newcommand{\bZ}{\mathbb{Z}}
\newcommand{\bR}{\mathbb{R}}
\newcommand{\cN}{\mathcal{N}}

\newcommand{\aaa}{\alpha}
\newcommand{\bbb}{\beta}

\begin{document}

\title{Hydrodynamic behaviour of an Abelian Sandpile Model with
  Laplacian rules}

\author{Andrea Sportiello}
\affiliation{Dip.~Fisica,
  Universit\`a degli Studi di Milano, and INFN, via G.\;Celoria 16,
  20133 Milano, Italy,}
\affiliation{LIPN, Universit\'e Paris-Nord, 
CNRS UMR7030,
99, av.\;J.-B.\;Cl\'{e}ment, 93430 Villetaneuse, France}
\email{Andrea.Sportiello@mi.infn.it}

\date{July 24th 2012}

\begin{abstract}
\noindent
We present a sandpile model, in which the instability of a site is
determined also by the variables in a neighbourhood. This is a
modification of the Abelian Sandpile Model, in which abelianity is
preserved: it shares several mathematical properties of the original
abelian model, while producing a more realistic dynamics. We show how
our model presents interesting hydrodynamic features.
\end{abstract}

\maketitle

\section{Introduction}
\label{sec.intro}

\noindent
Avalanche processes describe a variety of phenomena in Nature, ranging
from, of course, avalanches (e.g.\ in piles of sand), to epidemic
spread, allometric growth,\,\ldots\ They are considered as a possible
theoretical basis of $1/f$ noise (pink noise), and of the appearence
of Self-Organised Criticality in physical and natural
systems~\cite{BTW, CD}.

A lattice modelisation of non-equilibrium avalanche dynamics is
expected to be based on local \emph{toppling} and \emph{diffusion}
rules: if a toppling threshold inequality is satisfied at $x$ (i.e.,
$x$ is \emph{unstable}, e.g., the slope at $x$ is too steep), a
diffusion process occurs at $x$, which may trigger further topplings,
and so on, producing an avalanche.

When more sites are simultaneously unstable, an ordering prescription
must be given for definiteness. This produces a variety of (stochastic
and deterministic) possible rules, that share the same crucial
classification difficulties of non-equilibrium Statistical Mechanics,
in contrast with the more clear and rigorous understanding of
universality in equilibrium Critical Phenomena.

A simple dynamics, introduced by Bak, Tang and Wiesenfeld \cite{BTW},
involves a \emph{fixed}, and \emph{ultra-local} threshold rule, of the
form $z(x) > h$ (instead of a \emph{Laplacian} one, that involves
heights $z(y)$ at neighbours $y \sim x$). Remarkably, as shown in the
work of Dhar and collaborators \cite{Dhar, DharAlg}, for such a
dynamics topplings at different sites \emph{commute}, this justifying
the name of \emph{Abelian Sandpile Model} (ASM) given to this class of
systems. The ingredient of abelianity is crucial in the determination
of several hidden mathematical features of the model, among which,
notably, a bijection with \emph{spanning trees}, configurations in an
\emph{equilibrium} Statistical Mechanics model. Abelianity also allows
to achieve scaling to the continuum limit, either by the forementioned
correspondence,
or by the fact that adding particles and relaxing in rounds is
equivalent to add all the particles at once, and perform a unique
complete relaxation, this latter process being easier to analyse
theoretically in the large-volume limit (besides that algorithmically
convenient).

Unfortunately, abelianity seems to be a fragile ingredient. Small
modifications of threshold or diffusion rules easily break
commutativity. In particular, except for a construction specific to
1-dimensional chains, it seems impossible to implement genuine and
realistic toppling rules, based on gradient or Laplacian, within
ASM's. In \cite{usCipro} we presented a new class of ASM's, allowing
for multi-site topplings: this increases the spectrum of
possibilities, but still within ultra-local threshold rules.

Here we present a family of ASM's based on \emph{Laplacian-like}
toppling rules, possibly \emph{non-linear}.  The most natural rule
should be linear homogeneous, e.g.\ of the form 
$z(x) > \eval{z(y)}_{y \sim x}$. However we need to introduce
non-linearity, both for preserving abelianity, and for having a
compact space of configurations. The simplest variant is 
of the form $z(x) > \alpha \eval{z(y)}_{y \sim x} + \delta$,
thus linear inhomogeneous.
Other variants are discussed later on.

Furthermore, in order to preserve abelianity, we
need
to adopt
non-compact diffusion rules:
if ordinary nearest-neighbour
diffusion is $(D f)(x) = \sum_{y \sim x} \big( f(y)-f(x) \big)$, we
need a more general diffusion operator, of the form $(D_u f)(x) =
\sum_{y} u(x-y) \big( f(y)-f(x) \big)$. Fortunately, finite-range
functions $u(r)$, e.g.\ $u(r) \sim \exp(-\lambda |r|)$, are allowed, and
should show the same phenomenology of nearest-neighbour ones.


Interestingly, even in our largest class of models, configurations
recurrent under a dynamics of random increases of the
heights (plus relaxation), as in ordinary ASM's, still have an uniform
steady-state probability distribution, and a generating function given
by the Kirchhoff matrix-tree formula. It is possible that natural
bijections exist with spanning trees also in our wider context.

Thus we have continuous-variable sandpiles with fixed-threshold (F) or
non-local Laplacian-like (L) rules, and with compact (C) or non-compact
(N) diffusion. Of the four possibilities, the three (FC), (FN) and
(LN) have abelian realisations, with (FC) being the ``ordinary'' case.
We will use the shortcuts X-, Y-, XY-ASM (X=F,\,L; Y=C,\,N)
for these classes of models,
and ASM for the original model with discrete variables.

In order to support our claim that Laplacian-like threshold rules are
the crucial new ingredient here (while continuous variables and
non-compact diffusion are a technical accident), we show how
LN-ASM models have interesting
phenomenological properties
when observed under ``hydrodynamic'' experiments, that are not shared
neither with the ordinary
ASM and FC-ASM, nor with the
FN-ASM models.

\section{The model}
\label{sec.model}

\noindent
For simplicity we will only analyse translationally invariant
sandpiles with continuum height variables, on a portion 
$V = \bigotimes_{1 \leq \aaa \leq d} \bZ_{L_{\aaa}}$ of the
$d$-dimensional hypercubic lattice, with periodic or open boundary
conditions.  Extensions, to be discussed elsewhere, could include
models defined on arbitrary graphs, even directed, and realisations
with discrete variables.

We start by considering dynamics in which the heights are
\emph{increased}, stochastically or deterministically, and avalanches
are possibly produced.  When we say that a set of configurations is
``left stable by the dynamics'', we mean by any dynamics of this form.
In particular, this excludes
the inverse-toppling dynamics discussed in~\cite{usCipro}.

We denote by $\widetilde{w}(\vxi)$ the Laplace transform of $w(\vx)$.
We say that a function 
$f(\vx) : \bZ^d\to \bR$
is \emph{symmetric} if it has the symmetries of the cubic lattice,
$f(x_1,\ldots,x_d) = f(\epsilon_1
x_{\sigma(1)}, \ldots, \epsilon_d x_{\sigma(d)})$, for $\sigma \in
\mathfrak{S}_d$ and $\epsilon \in \{\pm 1\}^d$.

Our sandpile model is determined by two non-negative symmetric
functions $w(\vx)$ and $u(\vx)$, with
$w(\vzero) = u(\vzero) = 0$,
a dissipation parameter $\mu > 0$, and two functions $f(z)$, $g(z)$,
strictly- and weakly-monotonic respectively, with $g(z)$ having a
finite Lipschitz constant $\ell$. We say that $\vx$ is \emph{unstable}
if
\be
\label{eq.thres}
f\big( z(\vx) \big) 
>
\sum_{\vy}
w(\vy) \; g\big( z(\vx+\vy) \big)
\ef.
\ee
In such a case, a toppling $t_{\vx}$ may occur at $\vx$, modifying the
height function as
\be
t_{\vx}
\ : \ 
z(\vy) \to
\left\{
\begin{array}{ll}
z(\vy) + u(\vy-\vx) & \vy \neq \vx \\
z(\vx) - (1+\mu) 
\sum_{\vy'} 
u(\vy') & \vy = \vx
\end{array}
\right.
\ee
Note that $\mu>0$ implies that $\sum_{\vx} z(\vx)$ strictly decreases
at each toppling.  A configuration $z(\vx)$ is \emph{stable} if no
$\vx$ is unstable.  We have the 
F-ASM if $f(z)=z$ and $g(z)=1$.  Furthermore, it is customary to take
$w(\vx)=u(\vx) = \delta_{|\vx|,1}$ the indicator functions on nearest
neighbours, this producing a FC-ASM. We have an obvious affine
covariance $z \to \gamma_1 z + \gamma_0$, with $\gamma_1>0$, and
invariance under multiplication of the threshold rule (\ref{eq.thres})
by a positive constant, that we exploit later on.

We require the sandpile to have three properties:
\begin{description}
\item[\hbox{[A]}] A positive cone $\Omega = \{ z \,:\, z(\vx) >
  z_{\rm min} \textrm{~for all~} \vx\}$ is left stable by the
  dynamics.
\item[\hbox{[B]}] The set of stable configurations within $\Omega$ has
  finite non-zero measure.
\item[\hbox{[C]}] Within $\Omega$, the topplings are abelian.
\end{description}
Call $C = (1+\mu) \sum_{\vy \neq 0} u(\vy)>0$ and 
$\widetilde{w}(\vzero) = \sum_{\vy} w(\vy)$.  
We can use the covariance to fix
$z_{\rm min}=0$, i.e.\ $\Omega = (\bR^+)^V$. In this
case, condition {\bf [A]} means that, if $z(\vy)> 0$ for all $\vy$,
and $\vx$ is unstable, then (\ref{eq.thres}) implies $z(\vx)> C$,
that is, using the monotonicity, 
$C \leq C' :=f^{-1}\big( g(0) \widetilde{w}(\vzero) \big)$.

One easily sees that the set of stable configurations has non-zero
measure, because $(0,C']^V$ is a subset. Suppose that the limit for
$h \to +\infty$ of $f(h)- g(h) \widetilde{w}(\vzero)$ exists. If it is
positive, there exists $h_{\rm max}$ such that $f(h)> g(h)
\widetilde{w}(\vzero)$ for all $h>h_{\rm max}$, and any configuration
in $\Omega$
with $\max_{\vy} z(\vy) > h_{\rm max}$ is unstable at least at the
position of the max, so stable configurations are contained within
$(0,h_{\rm max}]^V$ and {\bf [B]} is verified. If it is negative,
there exists $\eps>0$, and an unbounded set of $h$'s in
$\bR^+$, such that the cubes $h-\eps < z(\vy) < h$ are stable,
because
$f(h)\leq \big( g(h) -\eps \ell \big) \widetilde{w}(\vzero) \leq
g(h-\eps) \widetilde{w}(\vzero)$, and {\bf [B]} is not verified.
When the limit is equal to zero, or does not exist, the condition
needs to be analysed more deeply, and we don't do this here.



A sufficient condition for {\bf [C]}, using the result on the analysis
of {\bf [A]} and the Lipschitzianity of $g(z)$, is that, for all $\vx
\neq \vzero$ and all $z > C'$,
\be
f\big( z+u(\vx) \big) - f(z)
>
\ell \sum_{\vy \neq \vzero}
w(\vy) u(\vx-\vy)
\ef.
\ee
Assume that $f'(z) \geq \ell' > 0$ for all $z \geq C'$. Then we have
the condition, for all $\vx \neq \vzero$,
\be
\ell' \, u(\vx) \geq 
\ell \,
\big(w \ast u\big)(\vx)
\ef.
\ee
It is easy to see that, if $u(\vx) = u_0 \exp(-\lambda |\vx|)$, then
$\sup_{\vx \neq \vzero} \big( w \ast u \big)(\vx)/u(\vx) =
\widetilde{w}(\lambda \nh)$ for $\nh$ some unit vector. For such a
function $u$ the condition reads $\widetilde{w}(\lambda \nh) \leq
\ell'/\ell$.  From now on, we will only consider $u$'s of this form.
Clearly $\widetilde{w}(\lambda \nh) > \widetilde{w}(\vzero)$ for all
$\lambda>0$, and the difference goes to zero in the limit, thus some
value $\lambda>0$ satisfying the condition exists if and only if
$\widetilde{w}(\vzero) < \ell'/\ell$.

Use affine covariance to set $C=1$. We can eliminate
$\widetilde{w}(\vzero)$ to get the sufficient and necessary condition
on $f$ and $g$, to admit $w$, $\lambda$ producing a LN-ASM
\be
\frac{f(1)}{g(0)} < \min
\left(
\frac{\ell'}{\ell} ,
\lim_{z \to \infty} \frac{f(z)}{g(z)}
\right)
\ef.
\ee
Recall that 
$\ell' = \min_{z > C'} f'(z)$ and $\ell = \max_{z>0} g'(z)$.
One easily sees $f(M) \geq M \ell' + \mathcal{O}(1)$
and $g(M) \leq M \ell + \mathcal{O}(1)$, implying that the minimum
above is always realised on the first quantity, i.e.\ the condition
{\bf [C]} is always stronger than {\bf [B]}.
In the case of a linear threshold rule, $g(z)=g_0+z g_1$ and
$f(z)=f_0+z f_1$ (with $g_0, f_1 >0$ and $g_1 \geq 0$) we get the
constraint $(f_0+f_1)g_1<f_1 g_0$. If $f(z)$ is quadratic, with
$f_2>0$, we get analogously $(f_0+f_1+f_2)g_1<(f_1+2 f_2)g_0$.

We say that a realisation of sandpile model as above is \emph{tight}
if conditions {\bf [A]} and {\bf [C]} are satisfied in a tight
way. Assuming that $f'(z)$ is monotonic, that $g(z) = g_0 + z g_1$ is
linear, and setting $C=1$, this means that $w$ and $\lambda$ are set
to satisfy
\be
\frac{f(1)}{g_0} = 
\widetilde{w}(\vzero)
<
\widetilde{w}(\lambda \nh)
=
\frac{f'(1)}{g_1}
\ef.
\label{asmtight}
\ee
For $W \subseteq V$, define $\chi_W(\vx)=1$ if $\vx \in W$ and 0
otherwise.  It is easy to see that the frame identity $\idf$ of the
LN-ASM on $V$ associated to $(f,g,w,u)$ depends
only on $u$ and the domain $V$ (in particular, it is the same as in the
associated FN-ASM), and is given by $\idf(\vx)=-\big(\chi_V \ast u
\big)(\vx)$, the difference in height after one toppling has occurred
at each site.  Conversely, the recurrent identity $\idr$ depends also
on $w$, $f$ and~$g$.

\section{Multiple threshold rules}

\noindent
For fixed $w$ and $u$, we may have families of LN-ASM, which provide a
continuous deformation of the FN-ASM with threshold rule $z(\vx)\geq
h$, and thus $f(z)=z$ and $g(z)=h/\widetilde{w}(\vzero)$.

We may extend the formalism of the previous section to functions $f, g
: \bR \to \bR^s$, for $s \geq 1$, and, in (\ref{eq.thres}), replace
$>$ with $\succ$, the canonical partial ordering ($f \succeq g$
iff $f_{\aaa} \geq g_{\aaa}$ for all $1 \leq \aaa \leq s$).
This provides a different mechanism for continuous
deformations of F-ASM's: we can take $s \geq 2$, and keep the original
rule for one of the components.

In the analysis of properties {\bf [A]} and {\bf [C]}, the different
components have a non-trivial interplay in a unique respect. For
property {\bf [A]} we need $C \leq C'_{\aaa} :=f_{\aaa}^{-1}\big(
g_{\aaa}(0) \widetilde{w}(\vzero) \big)$, for all $\aaa$. Then, for
property {\bf [C]} we need $\widetilde{w}(\lambda \nh) \leq
\ell'_{\aaa}/\ell_{\aaa}$, for all $\aaa$. Apparently, these
constraints are factorised, but this is not the case. While
$\ell_{\aaa}$, the Lipschitz constant of $g_{\aaa}(z)$, is in fact
independent from the other components, $\ell'_{\aaa}$ is defined in
terms of the allowed range for unstable heights, more precisely
\be
\ell'_{\aaa}
:=
\inf_{z > \max_{\bbb} (C'_{\bbb})}
\big( f'_{\aaa}(z) \big)
\ef.
\ee
Say that the (one or more) indices realising the max of $C'_{\bbb}$
are the \emph{leading components}. For a valid realisation of the
sandpile in dimension $s \geq 2$, the restriction of the threshold
equation to a subset of components containing at least one leading
component still produces a valid realisation. 
If we require non-redundancy, we get the constraint
\begin{description}
\item{\bf [D]} For each $1 \leq \aaa \leq s$, there exists $z \in
  \Omega$ such that $f_{\bbb}(z(\vx)) - \sum_{\vy} w(\vy)
  g_{\bbb}(z(\vy)) <0$
  only for $\bbb=\aaa$.
\end{description}
This constraint is specially important if we have $s=2$, $\aaa=1$ is
the fixed-threshold constraint $z(\vx) > h$, and is also the only
leading component.  Our sandpile model is genuinely distinct from the
original version if and only if the condition is verified, and the
point at which this happens in a tight way is the starting point of
the family of continuous deformations.

\section{Analysis of examples in $\boldsymbol{d=}$1 and 2}

\noindent
The tight realisations of the sandpile on an infinite linear chain,
with $w(\vx) = w\, \delta_{|\vx|,1}$ and $u(\vx) = u A^{-|\vx|}$,
require $A>1$ and give
\begin{align}
\frac{A^2 + 1}{2 A} 
&= 
\frac{f'(1) g_0}{f(1) g_1}
\ef,
&
u&=\frac{2}{A-1}
\ef,
&
w&=\frac{f(1)}{2g_0}
\ef.
\end{align}
If also $f(z)$ is linear, then we can set
\begin{align}
f(z) &=
z - 
\frac{(A-1)^2}{A^2 + 1}
\ef,
&
g(z) &= z+1
\ef,
\\
u&=\frac{2}{A-1}
\ef,
&
w&=\frac{2A}{A^2+1}
\ef.
\end{align}
The maximal possible height is $h_{\rm max}=(A^2+1)/(A-1)^{2}$.

\begin{figure}[tb!]
\setlength{\unitlength}{20pt}
\begin{picture}(5.6,5.25)(-0.3,-.5)
\put(0,4){\makebox[0pt][r]{\scriptsize{$40$}}}
\put(0,3){\makebox[0pt][r]{\scriptsize{$20$}}}
\put(0,2){\makebox[0pt][r]{\scriptsize{$0$}}}
\put(0,1){\makebox[0pt][r]{\scriptsize{$-20$}}}
\put(0,0){\makebox[0pt][r]{\scriptsize{$-40$}}}
\put(-.6,3.55){$x$}
\put(0.1,-.5){\makebox[0pt][c]{\scriptsize{$0$}}}
\put(1.1,-.5){\makebox[0pt][c]{\scriptsize{$20$}}}
\put(2.1,-.5){\makebox[0pt][c]{\scriptsize{$40$}}}
\put(3.1,-.5){\makebox[0pt][c]{\scriptsize{$60$}}}
\put(4.1,-.5){\makebox[0pt][c]{\scriptsize{$80$}}}
\put(5.1,-.5){\makebox[0pt][c]{\scriptsize{$100$}}}
\put(5.6,-.1){\makebox[0pt][c]{$t$}}
\put(0,0){\includegraphics[scale=1]{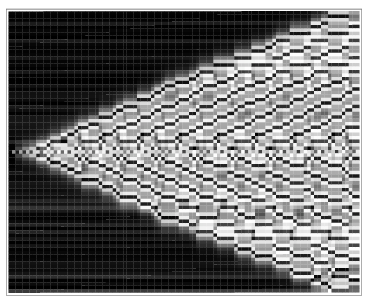}}
\end{picture}
\quad
\begin{picture}(5.6,5.25)(-0.3,-.5)
%
\put(0.1,-.5){\makebox[0pt][c]{\scriptsize{$0$}}}
\put(1.1,-.5){\makebox[0pt][c]{\scriptsize{$20$}}}
\put(2.1,-.5){\makebox[0pt][c]{\scriptsize{$40$}}}
\put(3.1,-.5){\makebox[0pt][c]{\scriptsize{$60$}}}
\put(4.1,-.5){\makebox[0pt][c]{\scriptsize{$80$}}}
\put(5.1,-.5){\makebox[0pt][c]{\scriptsize{$100$}}}
\put(5.6,-.1){\makebox[0pt][c]{$t$}}
\put(0,0){\includegraphics[scale=1]{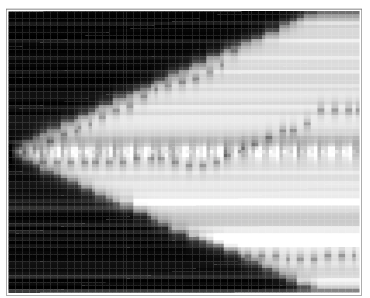}}
\end{picture}
\caption{\label{fig.1d}Time evolution of a centrally-seeded sandpile
  in $d=1$, with random low-height initialisation.  Right: in a LN-ASM
  as described in Setion~IV, with $A=2$. Left: in the associated
  FN-ASM.  In all our figures, black/white\,=\,low/high.}
\end{figure}

Figure \ref{fig.1d} compares the time trace of a simple protocol
(steady injection of sand in the middle of the lattice), between the
model above, for $A=2$, and the FN-ASM
using the same function $u(\vx)$. In both cases we observe a linear
growth of a plateau. While in the F-ASM the heights in the plateau
show fuzzy short-range fluctuations, in
the L-ASM the profile is smooth, with no need of coarse-graining.

\begin{figure}[b!]
\setlength{\unitlength}{23pt}
\begin{picture}(5,5.35)(-0.3,-.4)
\put(0,4.1){\makebox[0pt][r]{\scriptsize{$20$}}}
\put(0,3.1){\makebox[0pt][r]{\scriptsize{$10$}}}
\put(0,2.1){\makebox[0pt][r]{\scriptsize{$0$}}}
\put(0,1.1){\makebox[0pt][r]{\scriptsize{$-10$}}}
\put(0,0.1){\makebox[0pt][r]{\scriptsize{$-20$}}}
\put(2.25,-.4){\makebox[0pt][c]{\scriptsize{$0$}}}
\put(1.25,-.4){\makebox[0pt][c]{\scriptsize{$-10$}}}
\put(0.25,-.4){\makebox[0pt][c]{\scriptsize{$-20$}}}
\put(3.25,-.4){\makebox[0pt][c]{\scriptsize{$10$}}}
\put(4.25,-.4){\makebox[0pt][c]{\scriptsize{$20$}}}
\put(0,0){\includegraphics[scale=1.15]{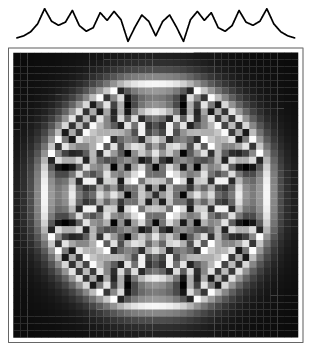}}
\end{picture}
\quad
\begin{picture}(5,5.35)(-0.3,-.4)
\put(0,4.1){\makebox[0pt][r]{\scriptsize{$20$}}}
\put(0,3.1){\makebox[0pt][r]{\scriptsize{$10$}}}
\put(0,2.1){\makebox[0pt][r]{\scriptsize{$0$}}}
\put(0,1.1){\makebox[0pt][r]{\scriptsize{$-10$}}}
\put(0,0.1){\makebox[0pt][r]{\scriptsize{$-20$}}}
\put(2.25,-.4){\makebox[0pt][c]{\scriptsize{$0$}}}
\put(1.25,-.4){\makebox[0pt][c]{\scriptsize{$-10$}}}
\put(0.25,-.4){\makebox[0pt][c]{\scriptsize{$-20$}}}
\put(3.25,-.4){\makebox[0pt][c]{\scriptsize{$10$}}}
\put(4.25,-.4){\makebox[0pt][c]{\scriptsize{$20$}}}
\put(0,0){\includegraphics[scale=1.15]{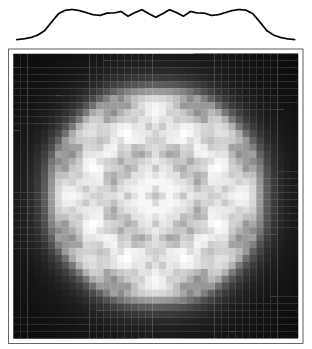}}
\end{picture}
\caption{\label{fig.2d}Relaxation of
  $z(\vx)=32/(1+|\vx|^2)$ in the FN-ASM (left),
  and of $z(\vx)=196/(1+|\vx|^2)$ in the LN-ASM (right).
On top: the profile of the middle row.
}
\end{figure}

In the two-dimensional case, we can still choose the model to be
tight, $w(\vx)=w \delta_{|\vx|,1}$, $g(z)=z+1$, $f(z)=z+f_0$,
$u(\vx)=u \exp(-\lambda |\vx|)$, and $C=C'=1$ (thus $f_0<0$,
$w=(1+f_0)/4$ and $\cosh(\lambda)=\frac{2}{w}-1$). As the series
$\cN(\lambda):=\sum_{\vx \neq \vzero} \exp(-\lambda |\vx|)$ has no
closed form, the parameter $u$ has to be set numerically to
$u=\cN(\lambda)^{-1}$.  For
example, choosing $f_0=-0.15$ gives
$u=0.1164$...\ This is the realisation presented in the example of
Figure~\ref{fig.2d}, and later on in the following section. Again, we
observe smoothness properties only in the LN-ASM case.

\section{Extreme regimes}

\begin{figure}[t!]
\makebox[0pt][c]
{
\includegraphics[scale=1.9]{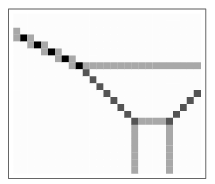}
\quad
\includegraphics[scale=1.9]{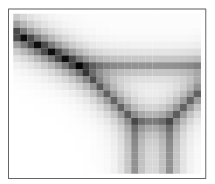}
\;
}
\caption{\label{fig.strings}Left: a portion of a typical outcome of
  toppling-antitoppling dynamics in the ASM. Right: the analogous
  outcome in our LN-ASM.}
\end{figure}

\noindent
We have seen above how the sandpiles with Laplacian-like 
toppling rules, when investigated
in regimes imitating realistic experimental settings (``hydrodynamic
regimes''), show smoothness properties not shared with the ordinary
versions. These settings are characterised by the presence of empty
regions, collecting the outcome of avalanches, and making them not
percolating.  In this section we discuss how the system behaves when
it is driven towards extreme regimes, in which the avalanche size is
regularised only by the finiteness of the volume.  In this case, the
new model preserves some resemblance with the intriguing combinatorial
features of the ordinary ASM.

A natural candidate is the study of the 
\emph{recurrent identity}, e.g.\ on a $L \times L$ portion of the
square lattice, with open boundary condition. This configuration has a
central role in the mathematical analysis of the model (it is the
identity of the abelian group associated to recurrent configurations)
\cite{Creutz}. It is obtained through a non-local procedure, as the
fixed point of iterated addition of frame identities, and is very
sensitive to the shape of the domain and boundary conditions.  In the
ordinary discrete ASM, it shows convergence to self-similar limit
shapes, involving fractal Sierpinskij-like structures.  An appropriate
discussion is too complex to fit here, and we start instead with
analysing its ``building blocks'', the \emph{strings} \cite{usEPL}.
These are one-dimensional structures in ASM's on bidimensional
lattices, that also play a role in the classification of the emerging
periodic patterns under deterministic protocols~\cite{Ostojic, DSC}.

In the ordinary (discrete) ASM, strings first appear when performing
``toppling--antitoppling'' operations \cite{usCipro, usEPL} on the
maximally-filled configuration
($z^{\rm (max)}(\vx)=3$ for all $\vx$, on the square lattice).  For
LN-ASM's we first need to generalise the concept of 
$z^{\rm (max)}(\vx)$, which turns out to be non-homogeneous.  It is
characterised as the unique solution to the system in which the
threshold inequalities (\ref{eq.thres}) are replaced by equalities.
Thus, it is stable, but $z^{\rm (max)}+\epsilon \,\chi_{ \{\vx\} }$ is
unstable for all $\vx \in V$, $\eps>0$.

If $f$, $g$ are linear and $w$ is nearest-neighbour, the corresponding
system has the form $(\Delta+a) z^{\rm (max)}(\vx) = b(\vx)$, plus
vanishing boundary conditions. In our geometry, this is easily solved
in Fourier basis and by method of images: the solution is smooth, and
near to $h_{\rm max}$, the maximal possible height, for sites far from
the boundary.

Next, we need to define toppling-antitoppling operators $\nu_{\vx}$ in
continuum ASM's.
Call $z'$ the relaxation of $z + s \chi_{\{\vx\}}$,
where $s$ is an amount making $\vx$ barely unstable. Then call $z''$
the antirelaxation of $z' - s \chi_{\{\vx\}}$. We set $\nu_{\vx} z :=
z''$.
The action of continuous toppling-antitoppling operators $\nu_{\vx}$
on $z^{\rm (max)}$, in LN-ASM, is analogous to the action of discrete
toppling-antitoppling operators $a^{\dagger}_{\vx}
a^{\phantom{\dagger}}_{\vx}$ in ASM. Examples are shown in
Figure~\ref{fig.strings}. 

The reason is that strings correspond to discontinuities in
piecewise-linear toppling matrices, and even in continuous sandpiles
the toppling matrices are discrete. This fact combines with the
observations in~\cite{DSC, usEPL}.

\begin{figure}[h!]
\makebox[0pt][c]{%
\includegraphics[scale=1.333]{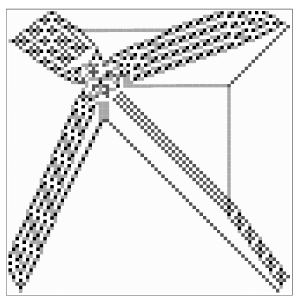}
\quad
\includegraphics[scale=1.333]{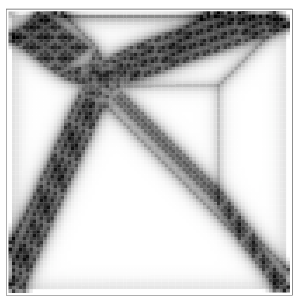}
}
\caption{\label{fig.prop81}Relaxing $\kappa L$ particles added at
  $\vx=(L/3,L/4)$ to $z(\vx)=h_{\rm max}$, on a $L \times L$ square
  (here $L=81$). Left: ASM, and $\kappa=2$. Right: GN-ASM, and
  $\kappa=4$. Graytones are scaled by $z^{\rm (max)}(\vx)$.}
\end{figure}

An aspect of this resemblance of ASM and LN-ASM occurs also in
a conceptually-simpler protocol: add sand 
to an uniform configuration $z(\vx)=h_{\rm max}$, 
at a unique site, 
then relax. Typical
outcomes are shown in Figure~\ref{fig.prop81}.

\vspace{1mm}
\noindent
{\bf Acknowledgements.}
Our work is supported by the
French ANR project MAGNUM
[BLAN-0204 07].



\vspace*{-4mm}



\end{document}